\documentclass[a4paper,fleqn,usenatbib]{mnras}

\usepackage{newtxtext,newtxmath}

\usepackage[T1]{fontenc}
\usepackage{ae,aecompl}

\usepackage[dvipsnames]{xcolor}

\usepackage{graphicx}	
\usepackage{amsmath}	
\usepackage{amssymb}	

\usepackage[normalem]{ulem} 
\usepackage{dsfont} 
\usepackage{subfigure}

\title[Resonant corrugation of a fast
  shock front]{Relativistic magnetohydrodynamical simulations of the resonant corrugation of a fast shock front}

\author[Demidem et al.]{
Camilia Demidem,$^{1,2}$\thanks{E-mail: demidem@apc.in2p3.fr}
Martin Lemoine,$^{2}$
Fabien Casse$^{1}$
\\
$^{1}$
Laboratoire AstroParticule \& Cosmologie (APC), Sorbonne Paris Cit\'e, Universit\'e Paris Diderot, CNRS/IN2P3, CEA/Irfu,\\ Observatoire de Paris, 10, rue Alice Domon et L\'eonie Duquet, F-75205 Paris Cedex 13, France\\
$^{2}$Institut d'Astrophysique de Paris, UMR 7095-CNRS, Universit\'e Pierre et Marie Curie - 98bis boulevard Arago, F-75014 Paris, France
}

\date{Accepted XXX. Received YYY; in original form ZZZ}

\pubyear{2017}

\begin{document}
\label{firstpage}
\pagerange{\pageref{firstpage}--\pageref{lastpage}}
\maketitle

\begin{abstract}
The generation of turbulence at magnetized shocks and its subsequent interaction with the latter is a key question of plasma- and high-energy astrophysics. 
This paper presents two-dimensional magnetohydrodynamic simulations of a fast shock front interacting with incoming upstream perturbations, described as harmonic entropy or fast magnetosonic waves, both in the relativistic and the sub-relativistic regimes. We discuss how the disturbances are transmitted into downstream turbulence and we compare the observed response for small amplitude waves to a recent linear calculation. In particular, we demonstrate the existence of a resonant response of the corrugation amplitude when the group velocity of the outgoing downstream fast mode matches the velocity of the shock front. We also present simulations of large amplitude waves to probe the non-linear regime. 

\end{abstract}




\section{Introduction}\label{sec:introd}
Collisionless shock waves are encountered in a wide variety of
astrophysical environments, on a wide range of flow velocities and
energy output, from our own solar system to supernova remnants and to
more extreme sources such as gamma ray bursts. In recent decades,
these phenomena have been receiving increasing attention, both from an
observational and from a theoretical perspective, all the more so with
the prospect of generating such shocks in the laboratory using
giant laser
facilities~\citep[e.g.][]{2012ApJ...749..171D,2017PhRvL.119b5001S}.

Collisionless shock waves appear as outstanding dissipation agents
and, near ubiquitously, as the sources of high energy particles and
non-thermal radiation. Although a detailed theoretical model of these
complex phenomena is still missing, our understanding has made
significant advances, thanks to the development of high performance
numerical simulations, in particular; see notably
\citet{2011A&ARv..19...42B}, \citet{2016RPPh...79d6901M} for recent
reviews.

Magnetohydrodynamical (MHD) turbulence proves to be an inseparable
feature of collisionless shock waves. It has long been recognized that
the generation of magnetized turbulence on plasma length scales is a
key element to structure the collisionless shock through collective
electromagnetic interactions~\citep[e.g.][]{1963JNuE....5...43M} and
to sustain the dissipation into a power-law of supra-thermal particles
\citep[e.g.][ and references therein]{1987PhR...154....1B}. How the
turbulence is generated and how it influences the shock physics are
thus two essential questions in this field of research.

Our present paper is connected to the latter question, and more
particularly to how MHD perturbations interact with a shock front, a
topic which itself possesses a rich literature, starting with
\citet{DIakov1958} and \citet{Kontorovich1958}. \citet{McKW71}, for
instance, have been interested in the possible amplification of
turbulence through shock crossing and on its phenomenological
consequences for the physics of the bow shock and the magnetopause;
ripples have indeed been observed in the Earth's bow shock, see
\citet{2006JGRA..111.9113M} and more recently
\citet{Johlander2016}. The possible amplification of turbulence in
spherical blast waves has also been suggested as a possible cause of
the ripples observed in some supernovae remnants -- see for instance
\citet{2008ApJ...689L.133B}, \citet{2011ApJ...735L..40B} and
\citet{2015ApJ...800...28Z} -- and how a shock, rippled by turbulence,
influences the physics of these objects has been discussed in a number
of studies, e.g. \citet{1986MNRAS.218..551A},
\citet{2001ApJ...563..800B}, \citet{2007ApJ...663L..41G},
\citet{2010ApJ...715..406G} or \citet{2012ApJ...747...98G}. More
recently, such interests have extended to the realm of relativistic
collisionless shock waves: \citet{2012MNRAS.422.3118L},
\citet{2016JPlPh..82d6301L}, and \citet{Zrake2016} have pointed out
the possible phenomenological consequences of the interaction of
turbulence with the termination shock of a pulsar wind, while
\cite{2007ApJ...671.1858S} and \citet{2011ApJ...734...77I} have been
interested in the relativistic generalization of the Richtmyer-Meshkov
instability at a corrugated shock front.

In this general context, \citet{LRG16} (hereafter LRG16) studied, in
the framework of linear perturbation theory, the stationary response
of a fast relativistic shock interacting with upstream or downstream
perturbations described as MHD harmonic waves (reminder in
Appendix~\ref{appendix:modes}).  This paper uncovered a resonant
response of the shock corrugation, i.e. a large or even formally
infinite deformation and transmission coefficient, for specific
characteristics of the incoming upstream wave. This process appears
reminiscent of the ``spontaneous emission of acoustic modes''
introduced by \citet{DIakov1958} and \citet{Kontorovich1958}, but LRG16
observed more precisely that the resonance occurs when the outgoing
fast magnetosonic mode -- namely, that transmitted in the downstream
plasma -- propagates with a group velocity that equals the shock
velocity; at such a resonance, the transmitted
  mode surfs on, and communicates its energy to the shock front.

The present work proposes to study this resonance through dedicated
MHD numerical simulations of the interaction of a harmonic mode with
a shock front. For a direct comparison to the results of the
previous study LRG16, we pay special attention to the case of
relativistic shock waves and conduct our simulations in
special-relativistic MHD (SRMHD); however, we will also show that
these results apply equally well to sub-relativistic shock waves so
that this resonance appears to be a universal phenomenon. Our MHD
simulations also allow us to study how this resonance evolves in the
non-linear regime, i.e. when perturbations of large amplitude interact
with the shock front.

In principle, corrugation can be induced by downstream fast MHD modes
outrunning the shock or by any kind of mode incoming from the
upstream. However, modes issued from far downstream can
  interact with the shock front only if their group velocity exceeds
  that of the shock, as viewed from the downstream rest frame. In this
  case, we further observe that the resonance takes place when the
  group velocity of the incoming downstream mode is very close to that
  of the shock front, so that 1) it formally takes a very long time
  for the incoming mode to catch up with the shock front, given that
  their relative velocity is small; 2) this resonance only appears on
  the boundary of the physical domain, i.e. the domain in which the
  incoming mode is able to catch up the shock front. We thus restrict
  our present analysis to the case of modes incoming from upstream. We further limit the study
to entropy and fast magnetosonic modes, without loss of generality as
the resonance phenomenon does not depend fundamentally on the nature
of the incident wave; we will provide more comments on this issue in
the following. Finally, we treat only 2D 
configurations, which are far less computationally expensive but still 
capture the essence of resonant corrugation.

This paper is organized as follows: Sect.~\ref{sec:2} briefly presents
the {\tt MPI-AMRVAC} code and our numerical setups, while our
results are reported in Sect.~\ref{sec:3}: the transfer
functions of an incoming entropy wave and a fast magnetosonic wave
interacting with a relativistic shock can be found in
Sect.~\ref{sec:3.1} while Sect.~\ref{sec:3.2} treats a
sub-relativistic case. Sect.~\ref{sec:4} outlines the main results and
provides some possible astrophysical implications.

\section{Relativistic planar MHD shock fronts}
\label{sec:2}
This section is devoted to the presentation of the physical framework used in our simulations. After briefly presenting 
 the equations governing such a formalism, we describe both the numerical methods as well as the setups used in our simulations.
\subsection{SRMHD framework}
\label{sec:2.1}

In this paper, we look at the temporal evolution of relativistic magnetized shock waves using a fluid approach, namely SRMHD. The governing equations of such description express the conservation of mass, momentum and energy density of the fluid. Simultaneously, it also provides 
the temporal evolution of the large-scale magnetic field including its interaction with the perfectly conducting fluid. Conservative equations read in CGS units as 
\begin{align}
   &\partial_t D +c \partial_j \left( D \beta^j\right) =
  0\,, \label{eq:continuity}\\[2mm] 
  &\partial_t S^i + c\partial_j
  \left\{ S^i \beta^j - \frac{1}{4\pi}\left[
  \frac{B^i }{\Gamma^2} +(\beta_k
  B^k ) \beta^i\right] B^j
  +P_{\rm tot}\delta^{ij} \right\} = 0\,,
    	\label{eq:momentum} \\[2mm]
    &\partial_t \tau +c \partial_j\left[ (\tau + P_{\rm tot}) \beta^j - \frac{1}{4\pi}(\beta_k B^k) B^j  \right] = 0\,,
    	\label{eq:energy}
\end{align} 
where the indices $(i,j,k)$ stand as $(x,y,z)$ components using the Einstein notation. The induction equation can be expressed thanks to 
the Ohm's law assuming a perfectly conducting fluid, namely 
\begin{equation}
\partial_t  B^i  + c\partial_j \left(  \beta^i B^j  -  B^i  \beta^j \right) = 0\,.
  \label{eq:faraday} 
\end{equation}
In the previous set of equations, $\beta^i=v^i/c$ is the component of the velocity along the $i$-direction normalized 
to the speed of light $c$. The associated Lorentz factor of the fluid is then $\Gamma=(1-\beta^2)^{-1/2}$ where $\beta^2=\beta_k\beta^k$. 
The mass energy density is
$D=\Gamma\rho c^2$ where $\rho$ is the proper mass density of the fluid. The relativistic momentum density along the $i$-direction is defined as 
$S^i=(\Gamma^2 w + B^2/4\pi) \beta^i-( \beta_k B^k) B^i/4\pi$ while the  total energy density of the fluid is denoted as $\tau=\Gamma^2 w+B^2/4\pi-P_{\mathrm{tot}}$ and  the magnetic field component along the $i$-direction as $B^i$. Finally, the quantity $w$ stands for the proper enthalpy density of the plasma and $P_{\mathrm{tot}}=p_{\rm th}+\left[B^2/\Gamma^2+(\beta_kB^k)^2\right]/8\pi$ is the total pressure associated with the thermal pressure $p_{\rm th}$ and the electromagnetic pressure.

In order to close the set of SRMHD equations, an equation of state (EOS) linking the thermal pressure to the enthalpy of the plasma has to be included. Following \citet{Meliani04} and \citet{Mignone2007}, 
   we can derive such a relation by considering the properties of the distribution function of a relativistic gas \citep{Taub48,Mathews1971}. This leads to  the following expression for the enthalpy:
  \begin{equation}
	w = \frac{5}{2} p_{\rm th} +\sqrt{\displaystyle \frac{9}{4}p_{\rm th}^2+\rho^2 c^4}\,. \nonumber
  \label{eq:enthalpy}
   \end{equation}   
  Or, equivalently, introducing the internal energy density $u$, so that $ w=\rho c^2+u+ p_{\rm th}$,
  \begin{equation}
  p_{\rm th} = \frac{u+2\rho c^2}{u+\rho c^2}\frac{u}{3}\,.
  \label{eq:eos}
   \end{equation}  
 It is noteworthy that the equivalent adiabatic index
$\gamma_\mathrm{eq} \equiv p_{\rm th}/u+1$ obtained with this
equation of state only differs by a few percents from that of the
theoretical Synge equation \citep{Synge1957,Mathews1971}.
 \subsection{Numerical methods}
The MPI-parallelized, Adaptive Mesh Refinement Versatile Advection Code ({\tt MPI-AMRVAC}) is a multi-dimensional numerical tool devoted to solve conservative equations  using finite volume techniques and a dynamically refined grid \citep{Holst2008,Keppens2012}. The {\tt MPI-AMRVAC} package handles hydrodynamical or magnetohydrodynamical equations either in a classical or relativistic framework. For the simulations displayed in this paper, we used a second-order Total Vanishing Diminishing Lax-Friedrichs (TVDLF) solver linked to a minmod slope limiter to make sure we employ a robust  scheme preventing any overestimate of the  corrugation of the shock front. 

The base level of the computational domain is filled with blocks of equal size, which can be divided into $2^D$ child grids having the same amount of grid cells than the parent grid ($D$ being the dimension of the grid). The structure of the grid will then be similar to an octree for three dimensional calculations. 
 The AMR refinement strategy can be controlled by several means within the {\tt MPI-AMRVAC} framework,  such as by Richardson extrapolation to future solutions or using instantaneous quantifications of the normalized second derivatives, or by a user controlled criterion or actually both \citep{Keppens2012}. For the purpose of our simulations, we simply choose to enforce the maximal refinement around the shock front and in the upstream in order to accurately describe the incoming wave and the corrugation of the shock front. 
 
 A potential downside of the finite-volume approach is that it does not guarantee that the magnetic field remains divergence free. This is of particular concern when considering highly magnetized relativistic shocks and indeed, we observe that without a method to correct the magnetic monopoles, unphysical errors (such as velocities larger than c) occur shortly after the incoming wave has encountered the shock. In order to overcome this problem, we have implemented within the {\tt MPI-AMRVAC} code a Constrained Transport algorithm based on \citet{Balsara1999}. In such approach, numerical fluxes provided by the SRMHD solver are used to enforce the solenoidal nature of the magnetic field (see Appendix~\ref{appendix:ct} for a comparison of the Constrained Transport method we used and the GLM divergence cleaning method, tested against the Orszag-Tang vortex problem).
 
\subsection{Initial set up and boundary conditions}
\label{sec:2.2} 
\begin{table*}
	\centering
	\begin{tabular}{lccccccccccr} 
		\hline
		setup & $\sigma_1$ & $\beta_{\rm pl,1}$ & $p_\mathrm{th,1}/\rho_1 c^2$ & $\beta_{1}$ & $\Gamma_1$ & $\rho_2 / \rho_1$ & $p_\mathrm{th,2}/\rho_2 c^2$ & $\beta_{2}$ & $\Gamma_2$ & $B_{2}/ B_{1}$ & mode\\
		\hline
		1 & 0.0996 & 0.02 & 0.001 & -0.9995 & 31.31 &   67.5 & 5.90 & -0.4206 & 1.102 & 2.38 & E\\
		2 & 0.1 & 1.59 &  $0.100$ & -0.9995 & 31.80 &  66.1 & 7.84 & -0.4334 & 1.110  & 2.31 & F\\
		3 & $1.0\times 10^{-4}$ & 0.2 & $1.0\times 10^{-5}$ & -0.1868 & 1.018 & 3.98 & $6.58\times 10^{-3}$ & -0.0477 & 1.001 & 3.92 & F\\
		\hline
	\end{tabular}
	\caption{Initial setup and nature of the incoming mode (entropy or fast mode) of our main simulations; indices $_1$
          (resp. $_2$) correspond to pre-shock (resp. post-shock)
          quantities. $\sigma_1\equiv {B_{1}}^2/\left(4\pi\Gamma_1^2
          w_1\right)$ represents the degree of magnetization of the
          upstream plasma, see text; we also indicate the
          plasma beta parameter $\beta_{\rm pl,1} \equiv 8\pi {\Gamma_1}^2
          p_{\mathrm{th},1}/{B_1}^2$ of the upstream medium.  The
          velocity and magnetic fields are evaluated in the shock rest
          frame. }
	\label{tab:setups}
\end{table*}
We initialize our simulations with the physical configuration 
of a stationary relativistic perpendicular shock exhibiting a
background magnetic field oriented along the $y$-direction while the upstream
plasma is flowing along the $x$-direction. The simulations are 2D
in the ($x$-$y$) plane, set in the shock rest frame and we express physical quantities 
in this frame, unless stated otherwise. The initial set up
then corresponds to the exact solution of the
Rankine-Hugoniot jump relations in the shock frame \cite[see][e.g]{Goedbloed2010}, namely 
\begin{align}
    &\rho_1 \Gamma_1 \beta_1 = \rho_2 \Gamma_2 \beta_2\,,
    \label{eq:cont }\\[2mm]
    &B_1 \beta_1  = B_2 \beta_2 \,,
    \label{eq:bflux }\\[2mm]
    &W_1 {\Gamma_1}^2 {\beta_1}^2 +P_{\rm tot,1} = W_2 {\Gamma_2}^2{ \beta_2}^2 + P_{\rm tot,2}\,, 
    \label{eq:ei1}\\[2mm]
   &W_1 {\Gamma_1}^2 \beta_1 = W_2 {\Gamma_2}^2 \beta_2\,, 
   \label{eq:ei2}
\end{align} 
where upstream and downstream quantities are referred to, with the
index $_1$ and $_2$, respectively.
Again, $P_{\rm tot}$ represents the
generalized pressure, which now reads $P_{\rm tot}=p_\mathrm{th} +
B^2/(8\pi\Gamma^2)$ and $W$ the generalized proper enthalpy density,
$W=w+B^2/(4\pi\Gamma^2)$. We quantify the degree of magnetization of
the upstream through the magnetization parameter:
$\sigma_1\,\equiv\,B_1^2/\left(4\pi{\Gamma_1}^2w_1\right)$, which is
equivalently related to the proper Alfv\'en 3-velocity $\beta_{\rm A,1}$ of the
upstream plasma through $\beta_{\rm A,1}^2=\sigma_1/(1+\sigma_1)$
since $B_1/\Gamma_1$ corresponds to the magnetic field strength in the
proper upstream frame. The initial setup and the nature of the mode imposed at the inflow boundary of the simulations that we discuss in this paper are summarized in Table~\ref{tab:setups}. For convenience, we also indicate there the plasma beta parameter $\beta_{\rm pl,1} \equiv 8\pi {\Gamma_1}^2 p_\mathrm{th,1}/{B_1}^2$.
The value of the adiabatic index is not crucial for the problem at hand, but we adopted an EOS in agreement with Eq.~\ref{eq:eos}, which gives, for both sides of the rippling shock, values more realistic and closer to the analytic study we compare our results with; for the relativistic simulations: $\approx5/3$ and $\approx4/3$ in the upstream and downstream medium respectively, and $\approx5/3$ in the entire simulation box for sub-relativistic simulations.\footnote{In practice, the relative variations of the index, $\delta \gamma_\mathrm{eq} /\gamma_\mathrm{eq}$, are (at most) of the order of the percent. Simulations of setup 1 were run with a constant polytropic index of $4/3$ and show no noticeable difference with simulations run with a varying index, aside from the slightly different initial conditions.}

At the beginning of the simulation, we launch an incoming wave, whose characteristics match the relevant
analytical expressions of the desired linear MHD mode (see
Appendix~\ref{appendix:modes}), from the right (upstream) $x$-boundary
of the computational domain, then study the reaction of the shock
front over a timescale sufficient to see a stationary regime establishing
itself. The incoming wave is harmonic, either an entropy or a fast
magnetosonic mode, with a wavevector lying in the ($x$,$y$) plane, so that the problem remains 2D. As discussed in the following, such simulations
are rather time consuming because they require a high resolution in
order to observe the corrugation in the linear limit; we thus restrict
our study to a range of wavenumbers around the resonance brought
forward by the study of LRG16.

Although LRG16 contains some discussion about how the resonance arises, it will prove useful to explain this point in some detail. We determine this resonance through a numerical computation of the
longitudinal wavenumber $k_x$ of the incident mode, which is such that
the velocity of the outgoing fast magnetosonic mode matches the shock
front velocity, as follows.  In a linearized analysis, the incoming upstream perturbation is transmitted
through the corrugated shock front as a set of downstream MHD modes,
which pulsate at the same frequency $\omega$ as the incoming mode and
the corrugated shock front (in the shock rest frame). In the rest
frame of the downstream plasma, the frequency of these outgoing modes is Doppler boosted to
$\omega_2$, according to: $\omega\,=\,\Gamma_2\left(\omega_2+\beta_2c
k_{x,2}\right)$, where $k_{x,2}$ corresponds to the (mode-dependent)
$x-$wavenumber of the outgoing wave in the downstream rest frame. The value of this
wavenumber is determined by the dispersion relation of the
corresponding MHD mode, which relates $\omega_2$ to $k_{x,2}$, hence
$\omega$ to $k_{x,2}$ through the previous frequency matching; all
modes share of course the same perpendicular wavenumber $k_y$. In
turn, $\omega$ is directly related to the $x-$wavenumber of the
incoming mode through its own dispersion relation, therefore once the
perpendicular wavenumber is fixed, the longitudinal wavenumbers of the
outgoing modes are direct functions of the longitudinal wavenumber of
the incoming mode. 

Regarding the entropy mode, which is generically excited
  by the corrugation of the shock front, its dispersion relation in
  the downstream rest frame is $\omega_{2,\rm E}\,=\,0$, so that its
  $x-$wavenumber is $k_{x,2,\rm E}\,=\,\omega/(\Gamma_2\beta_2 c)$.
   
For the outgoing magnetosonic modes, the dispersion relation in the downstream rest frame takes the
form of a quartic equation:
\begin{equation}
\begin{split}
  \omega_2^4 - \left[\beta_{\rm F,2}^2(k_{x,2}^2 \right.+ k_y^2) &+ \left.\beta_{\rm A,2}^2 \beta_{\rm s,2}^2 k_y^2\right]c^2\omega_2^2 \\ 
      &+\beta_{\rm A,2}^2 \beta_{\rm
    s,2}^2c^4(k_{x,2}^2+k_y^2)k_y^2\,=\,0\,,
\end{split}
\end{equation}
where $\beta_{\rm A,2}$, $\beta_{\rm s,2}$ respectively denote the Alfv\'en
and sound velocities of the shocked plasma, while $\beta_{\rm
  F,2}^2\,\equiv\,\beta_{\rm A,2}^2+\beta_{\rm s,2}^2-\beta_{\rm A,2}^2 \beta_{\rm
  s,2}^2$. Solving this dispersion relation, one obtains 4 outgoing magnetosonic modes. For each mode, one can
compute the group velocity $c\beta_{\rm g,2}\,\equiv\,{\rm
  d}\omega_2/\rm{d}\mathrm k_2$ (as defined in the downstream
plasma rest frame). One then finds that there are always two outgoing slow modes
propagating slower than the shock front, relative to the downstream
plasma, as they should indeed for a fast shock. The two remaining solutions are either fast modes, one of which can be discarded as it outruns the shock, i.e. $\vert \beta_{{\rm
    g},2,x}\vert\,>\,\vert \beta_{2}\vert$ and $\beta_{{\rm
    g},2,x} \beta_{2}<0$ (since $-\beta_{2}$ corresponds to the shock velocity relative to the downstream) or, two waves with complex wavenumbers, one of which is unphysical, as it diverges far from the shock, while the other describe a surface wave on the front (see LRG16 for
further discussion on the number of degrees of freedom of the outgoing
modes, see also \citet{Lubchich2005} for a detailed
  discussion on the nature of the modes). The resonance emerges at the transition between these
two cases, when the velocity of the outgoing mode nearly coincides
with the shock velocity in the downstream plasma frame.

We used a fixed grid, uniform for simulations of large amplitude incoming waves and refined in the upstream and close post-shock regions for low amplitude waves. The resolution is mainly constrained by the amplitude of the corrugation in the $x$-direction. The simulations with the lowest number of cells were run on a uniform $1920\times120$ grid while the simulations with the highest number of cells were run on a $10560\times72$ base grid with 3 levels of refinement i.e. a local resolution in the upstream and shock regions 16 times larger.

The upper and lower $y$-boundaries were periodic and the left $x$-boundary ensured continuous fields and corresponded to the downstream outflow.
For the least corrugated shocks, the deformation could only be
resolved over a few cells, which entails some errors in the
measurement of the corrugation amplitude, but the incident wave was
always largely sampled (about a thousand cells per $x$-wavelength). We
checked that, at least for small amplitude waves -- i.e.,
perturbations of the order of the percent, $\delta \rho /\rho\sim 1\%$
-- the polarization of the wave when it reaches the shock coincides to
that we input at the border of the simulation box. For larger
amplitude ($\delta \rho /\rho\sim 1$) magnetosonic waves, some mode
conversion occurs during the propagation resulting into amplitude
changes of a few percents.

We do not observe any major influence of the resolution on the results
of Sec.~\ref{sec:3}: for increasing resolution, the measured amplitudes remain compatible with each other inside error bars of decreasing magnitude and the small scale structures are less dissipated.
 The size of the box in the $x$-direction does not
affect the results either, as long as it is larger than a few
transverse wavelengths of the incident perturbation $\lambda_y$. In
practice, we set the simulation box size so that the downstream length
is a few $x$-wavelengths of the largest scale outgoing mode. The
upstream extension has no influence but the larger it is, the more
mode conversion can develop for large amplitude waves, and the more
the wave is damped before reaching the shock due to numerical dissipation. Regarding the size of
the box in the $y$-direction, as long as it spans $n$ $\lambda_y$, the
same pattern is simply repeated $n$ times along the vertical.

\section{Results}
\label{sec:3}

\begin{figure*}
\center
\includegraphics[scale=0.43]{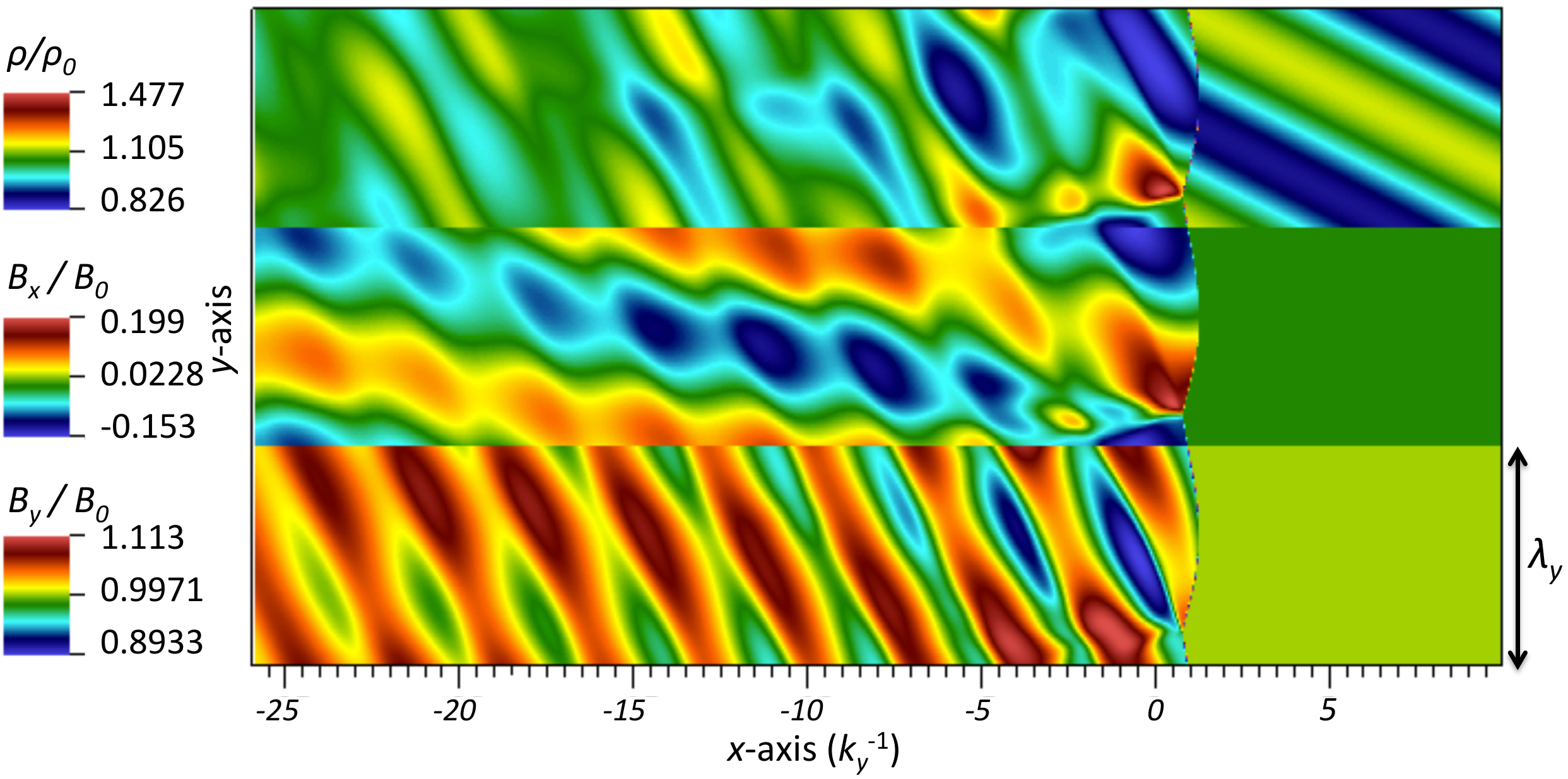}
\caption{Ratios of the density and magnetic field components to their initial values for an incident entropy wave of amplitude $\delta \rho / \rho\approx0.13$ and wavevector close to the resonance, interacting with the shock of setup 1, in a stationary regime. The transverse size of one panel is $\lambda_y \equiv 2\pi/k_y$. }
\label{fig:snapshot_e_a13}
\end{figure*}
\begin{figure*}
\center
\includegraphics[scale=0.43]{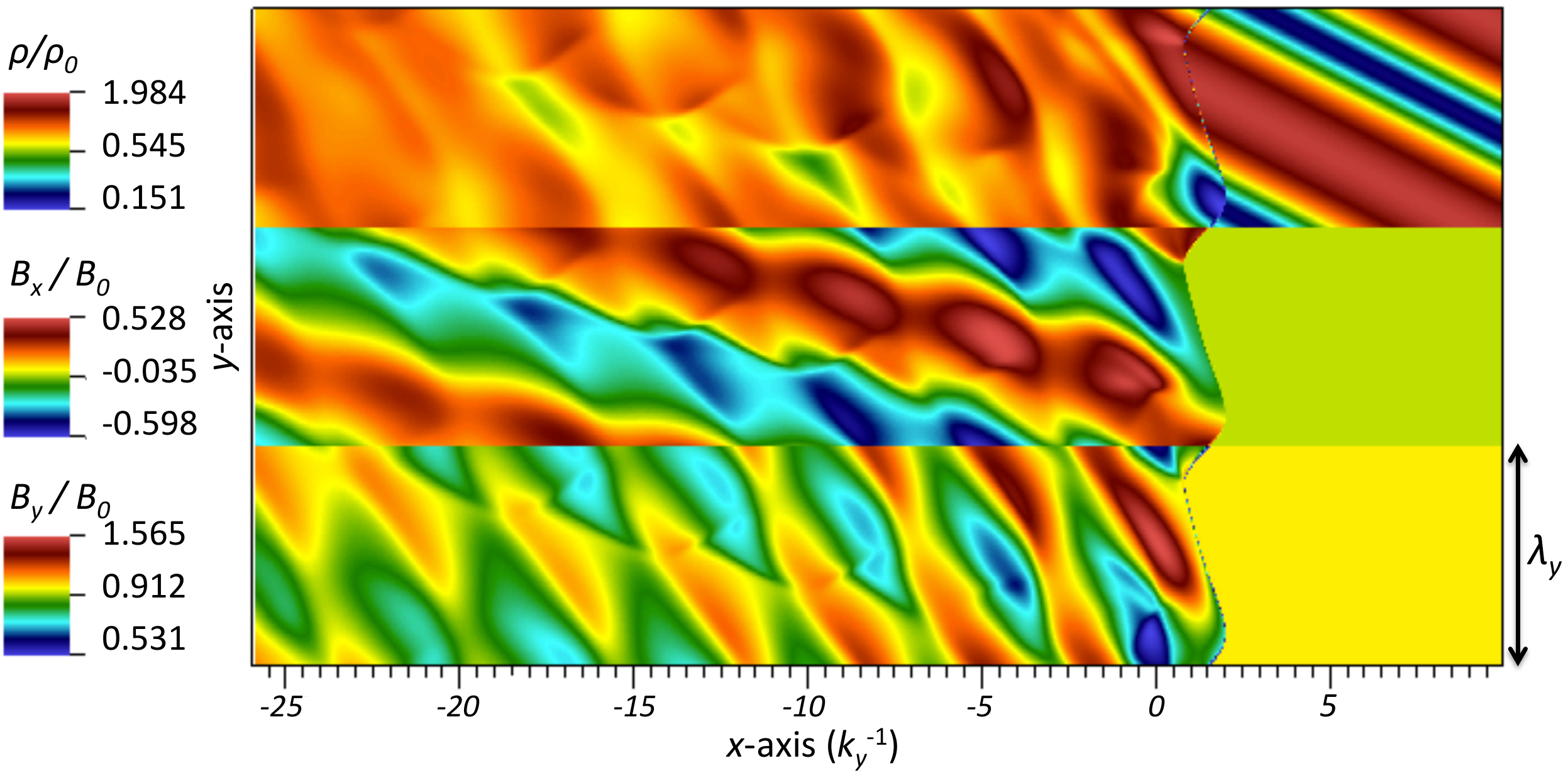}
\caption{Snapshot from a simulation with the same parameters as in Fig.~\ref{fig:snapshot_e_a13} but for an incident wave of larger amplitude $\delta \rho / \rho\approx0.8$.  }
\label{fig:snapshot_e_a80}
\end{figure*}

The typical timeline of a simulation is the following: as the incident
wave impinges the shock front, corrugation develops and downstream
modes are generated, then propagate away from the shock at their own
group velocities. Close to the resonance, for small amplitude incoming waves, the amplitude of the
corrugation increases slowly until it eventually reaches a stationary
state, typically over a timescale of a few hundreds of $\omega^{-1}$. For non-resonant and/or large amplitude incoming waves, the final corrugation amplitude is reached on timescales of a few $\sim \omega^{-1}$, with some fluctuations though, for large amplitude incoming waves with a wavevector close to the resonance.

Once the source is shut off, the shock slowly regains planarity. 

In the following, we first discuss the case of relativistic magnetized shock fronts,
to make contact with the linear theory developed in LRG16, then we
analyse the corrugation of sub-relativistic shock waves.

\subsection{Interaction of an upstream mode with a relativistic shock front}
\label{sec:3.1}

\begin{figure*}
\center \includegraphics[scale=0.43]{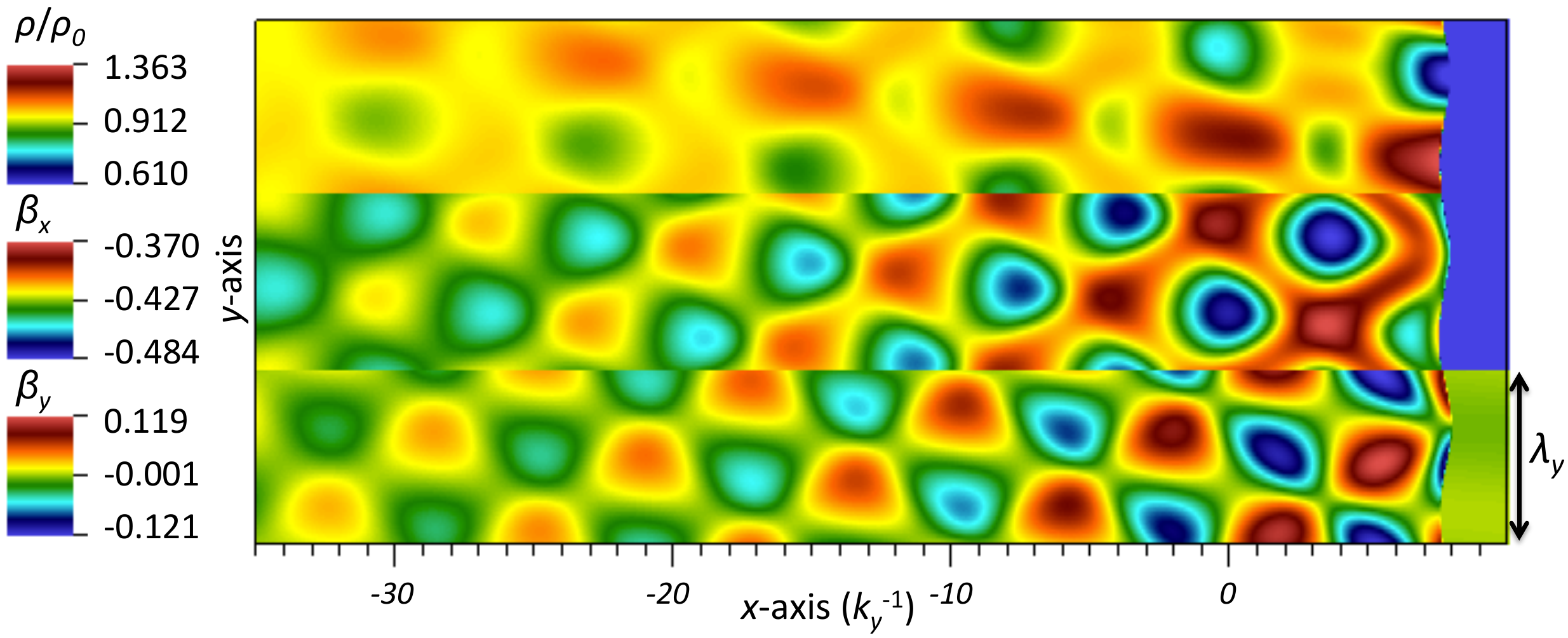}
\caption{Downstream density and velocity profiles for an incoming magnetosonic wave of
  amplitude in density $\delta \rho / \rho\approx0.45$ interacting with the
  shock of setup 2, at late times. $\rho_0=\rho_2$ is the initial downstream proper density and the color map was truncated to enhance the downstream structures.}
\label{fig:snapshot_ms}
\end{figure*}

We consider here the case of a relativistic shock wave (setups 1 and 2
of Table~\ref{tab:setups}) for which $\Gamma_1\,\approx\, 30$.

\subsubsection{Incoming entropy modes -- setup 1}

Fig.~\ref{fig:snapshot_e_a13} presents a snapshot of a simulation
corresponding to setup 1, for an incoming entropy wave, i.e. density perturbations, with $\delta
\rho / \rho\approx0.13$ and a wavevector close to the resonance.


For small amplitude waves, corresponding to the linear interaction
regime, the downstream medium can be described as a superposition of
MHD modes, namely an entropy mode and two slow magnetosonic modes,
plus a fast magnetosonic mode for wavevectors larger than the
resonant one. As expected, we observed no downstream Alfv\'en waves since the
specific geometry of these simulations is 2D. Indeed, setting the magnetic
field in the plane of the simulation eliminates transverse waves
(unless $k_x\,=\,0$, which would lead to degenerate Alfv\'en
modes). This is not a strong restriction, since the linear theoretical
analysis indicates that incoming compressible modes with a wavevector lying in the ($x$,$y$) plane are
converted into outgoing compressible modes. To confirm this, we
carried out dedicated 2.5D simulations, in which the spatial dependence of the physical fields is still 2D but where magnetic \& velocity vectors can have arbitrary orientations in 3D. Such configuration hence allows transverse modes, but we did not find any trace of Alfv\'en waves. Therefore, these waves are likely to play a role only in full 3D configurations or for the case of incoming Alfv\'en waves. Incidentally, we also note that taking
the background magnetic field along $z$ out of the simulation plane
makes very little difference: the downstream turbulence structure
appears somewhat simpler since for $\boldsymbol{k} \perp
\boldsymbol{B}$, only fast magnetosonic modes can propagate. 

Fig.~\ref{fig:snapshot_e_a80} shows a snapshot from a simulation with the same physical parameters and numerical resolution as in Fig.~\ref{fig:snapshot_e_a13}, but for a larger amplitude of the incoming wave, $\delta
\rho / \rho\approx0.8$. 
The various flow quantities in the downstream medium are
perturbed well into the non-linear regime, which leads to non-linear
interactions remodelling the flow away from the shock and to the 
dissipation of small-scale structures.
  
\subsubsection{Incoming fast magnetosonic modes -- setup 2}
\label{sec:3.1.2}
We consider here a similar set-up as in the previous case, but for an
incoming magnetosonic mode; in this case, all flow quantities of the
upstream medium are perturbed, (see Appendix~\ref{appendix:modes}). The simulations show qualitatively the same features
as for entropy waves, with the same turbulence pattern at similar wavelengths. Fig.~\ref{fig:snapshot_ms} shows an
example in the non-linear regime, $\delta
\rho/\rho\,\approx\,0.45$, away from the resonance (with $k_x$ smaller than the one giving rise to resonance). The
initial perturbation is not visible as we expressed the density in units of the initial downstream density and the scale of the colormap was truncated to enhance the downstream
structures. Here as well, we observe non-linear
turbulence, with typical mildly relativistic velocities, on the
downstream side, which is remodelled through non-linear interactions
in time, i.e. away from the shock. The dissipation we observe is mainly of numerical nature but could also be partly accounted for by destructive interference and by the surface mode with complex $k_x$ mentioned in Sec.~\ref{sec:2.2}.

\subsubsection{Transfer function}
\label{sec:3.1.3}

\begin{figure}
\center \includegraphics[scale=1.]{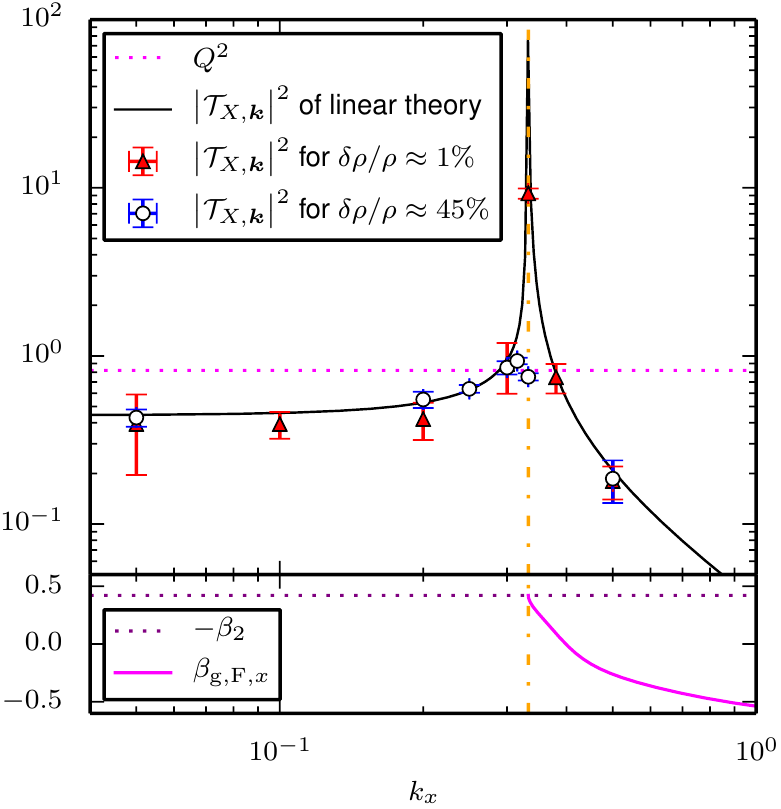}
\caption{Top panel: squared modulus of the transfer function
  $\mathcal{T}_{X,\boldsymbol{k}}$ for an incoming entropy wave as a
  function of $k_x$, for setup 1 and $k_y=1$ in arbitrary units. The
  transverse size of the box is $2\pi$ in these units. The solid black
  curve plots the prediction of the linear analysis of LRG16; the
  symbols correspond to the results of our MHD simulations, in
  triangles for a small-amplitude wave $\delta\rho/\rho\,\approx\,0.01$ and
  in circles for a large-amplitude wave
  $\delta\rho/\rho\,\approx\,0.45$. The error bars give the measurement
  uncertainty on the corrugation amplitude $\Delta X_{\boldsymbol{k}}$
  due to the finite resolution or to small non-stationarity for resonant large amplitude waves. The dotted line indicates the measure
  $Q^2$ of the perturbation of the upstream energy momentum tensor in
  units of $\delta\psi$.  In the bottom panel, we plot the
  corresponding (analytically computed) group velocity along $x$ of
  the outgoing downstream fast magnetosonic mode; it intersects the
  shock speed line at the resonant $k_x$, which confirms the origin of
  the resonance seen in the upper panel. All quantities are evaluated in the downstream rest-frame.}
\label{plot:tf_e}
\end{figure}

The induced corrugation can be quantified through a transfer function, 
\begin{equation}
\mathcal{T}_{X,\boldsymbol{k}} \,\equiv\, k_{y}\, \frac{\Delta X_{\boldsymbol{k}}} {\delta\psi}  ,
\label{eq: transfer_fct}
\end{equation}
which relates the amplitude $\delta\psi$ of the incoming wave displaying a wavevector
$\boldsymbol{k}=(k_{x},k_{y})$ and, the corrugation amplitude $\Delta X_{\boldsymbol{k}}$
(see Appendix~\ref{appendix:modes} for the definition of $\delta\psi$). The $k_{y}$ factor ensures
that $\mathcal{T}_{X,\boldsymbol{k}}$ is a dimensionless quantity; if the transverse wavelength
is increased, $\Delta X_{\boldsymbol{k}}$ will increase as much and
$\mathcal{T}_{X,\boldsymbol{k}}$ will remain unchanged.

The corrugation amplitude, $\Delta X_{\boldsymbol{k}}$, is the amplitude along the $x$-direction of the rippled shock. In a fluid approach, the shock theoretically corresponds to a discontinuity appearing in some physical quantities. In our simulations however, its width is finite because of numerical diffusion, but is still much smaller than the corrugation scale, provided the resolution is high enough. We can then, quite arbitrarily, materialize the shock location as the line of cells where one of these fields (the density or the Lorentz factor, for instance) is the average of the background upstream and downstream values. The corrugation amplitude is then simply half of the peak-to-peak amplitude of the deformation of this line.

The definition of an amplitude $\delta\psi$ for the incoming wave is also
somewhat arbitrary, because although the various flow variables all
scale linearly with $\delta\psi$ (see Appendix~\ref{appendix:modes}), they
do so differently in terms of $k_x$ and $k_y$.  This simultaneous
dependence on $\delta\psi$ and $k_x$ notably implies that, at fixed
$\delta\psi$, the perturbations of the components of the
energy-momentum tensor evolve in non-trivial (and different) ways in
terms of $k_x$. It is thus possible, in principle, to send in a
wave with $\delta\psi\,\ll\,1$ which corresponds to a large
perturbation of the energy or momentum flux along the shock normal,
and which leads to a large response of the shock front. In order to
distinguish such a response from a resonant response, we also keep
track of the following quantity:
\begin{equation}
  Q\,\equiv\,\frac{1}{\sqrt{2}\delta\psi}\left[\left(\frac{\delta T^{tx}}{
      T^{tx}}\right)^2+\left(\frac{\delta
      T^{xx}}{T^{xx}}\right)^2\right]^{1/2}
  \label{eq:dQ}
\end{equation}
where $\delta T^{\mu \nu}$ and $T^{\mu \nu}$ are the perturbed and
unperturbed energy-momentum tensors of the upstream plasma. $Q$ thus provides a measure of the perturbation in the incoming energy-momentum flux in units of $\delta \psi$ and the resonant response we are looking for is such that ${\mathcal{T}_{X,\boldsymbol{k}}}^2>Q^2$.

\begin{figure}
\center
\includegraphics[scale=1.]{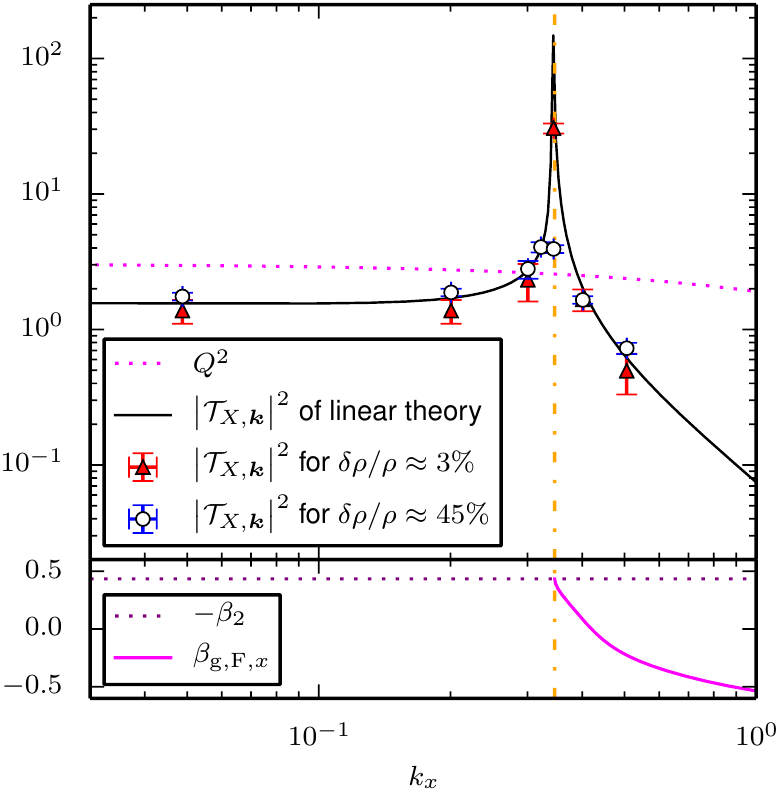}
\caption{Same as Fig.~\ref{plot:tf_e} for an incoming fast
  magnetosonic mode corresponding to setup 2 with $k_y=1$ (arbitrary
  units) and wave amplitudes as indicated. All quantities are evaluated in the downstream rest-frame.}
\label{plot:tf_ms}
\end{figure}

We compare here the transfer functions, obtained by solving numerically the shock crossing conditions
perturbed at the first order as presented in LRG16 (black curve) and
by measuring the shock deformation in the simulations (data points).
Figure~\ref{plot:tf_e} corresponds to the entropy wave of
Sect.~\ref{sec:3.1} (setup 1) and Fig.~\ref{plot:tf_ms} to the fast
magnetosonic mode of Sect.~\ref{sec:3.2} (setup 2). 
The error bars give the measurement
  uncertainty on the corrugation amplitude 
  due to the finite resolution and to some non-stationary features close to the resonance for large amplitude waves.
 The dotted lines shown in the upper panels
of these figures give the value of $Q^2$, for comparison to ${\mathcal{T}_{X,\boldsymbol{k}}}^2$ as discussed above. 
In Fig.~\ref{plot:tf_e}, this dependence is trivial $Q\,\approx\,1$, since for entropy modes, the perturbations are
independent of the wavevector, but in Fig.~\ref{plot:tf_ms}, $Q$ depends somewhat on $k_x$; such a dependence will
be exacerbated in the sub-relativistic case of setup 3 that we study next.

Both figures clearly reveal a resonant response of the shock
corrugation to the incoming perturbation, in good agreement with the
prediction of the linear theory. In each figure, the lower panel plots
the group velocity along $x$ of the outgoing fast magnetosonic mode as a
function of the incoming longitudinal wavenumber. These panels confirm
that the resonance occurs when the outgoing mode travels at a velocity
close to that of the shock front. For values of $k_x$ smaller than the
resonant one, the group velocity becomes complex, as the mode then
turns into a surface wave located on the shock front, see the
discussion above.

As an order of magnitude, for the small amplitude entropy wave simulations, for which
$\delta\psi\,=\,\delta \rho / \rho$, $k\Delta X_{\boldsymbol{k}}\approx
3.8\delta\psi$ close to the resonance while $k\Delta X_{\boldsymbol{k}}\approx
0.5\delta\psi$ at large $k_x$. Linear theory predicts a more
pronounced resonance with $k\Delta X_{\boldsymbol{k}}\,\gtrsim\, 10\delta\psi$.

\begin{figure}
\center
\includegraphics[scale=1.]{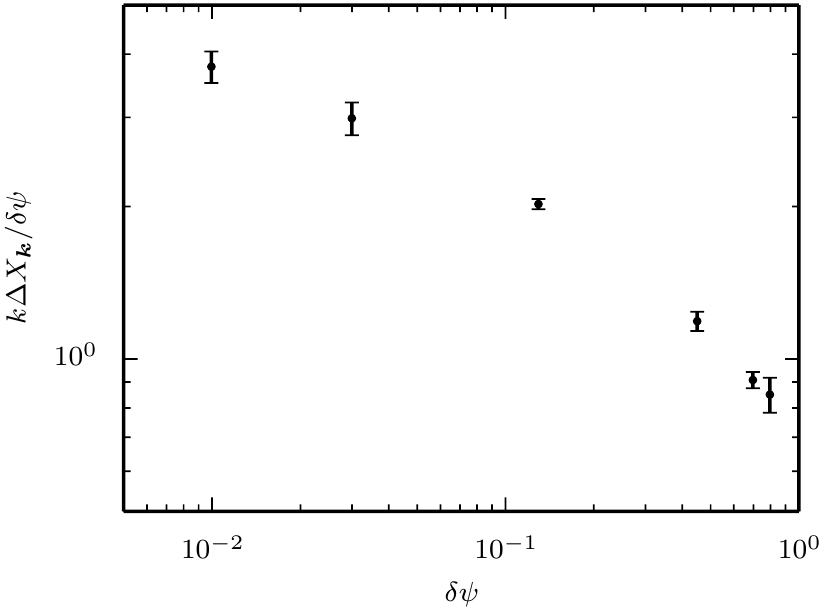}
\caption{$k\Delta X_{\boldsymbol{k}}/\delta \psi$ measured in the simulations of setup 1 close
to the resonance for incoming entropy waves of increasing amplitudes $\delta \psi=\delta \rho/\rho$. }
\label{plot:ampli}
\end{figure}

Figures~\ref{plot:tf_e} and \ref{plot:tf_ms} also plot the response of
the shock to perturbations of significant amplitude,
$\delta\rho/\rho\,\approx\,0.45$, beyond the reach of linear theory. We
observe that the resonance remains, but that it is smoothed
out with a typical response $k \Delta X_{\boldsymbol{k}}\,\approx\,1$, as anticipated
in \citet{2016JPlPh..82d6301L}. The observed smoothing is rather evocative of a
non-linear resonance broadening effect; it likely results from the
non-linear couplings of the MHD equations, which become sizable at
large amplitude, and which imply that the eigenmodes of linear MHD
have a finite lifetime against conversion. 
This scaling $k\Delta X_{\boldsymbol{k}}\,\approx\,1$ is compared in
Fig.~\ref{plot:ampli} against the simulations of setup 1 at various
wave amplitudes.

\subsection{Interaction with a sub-relativistic shock front}
\label{sec:3.2}

\begin{figure*}
\center
\includegraphics[scale=0.52]{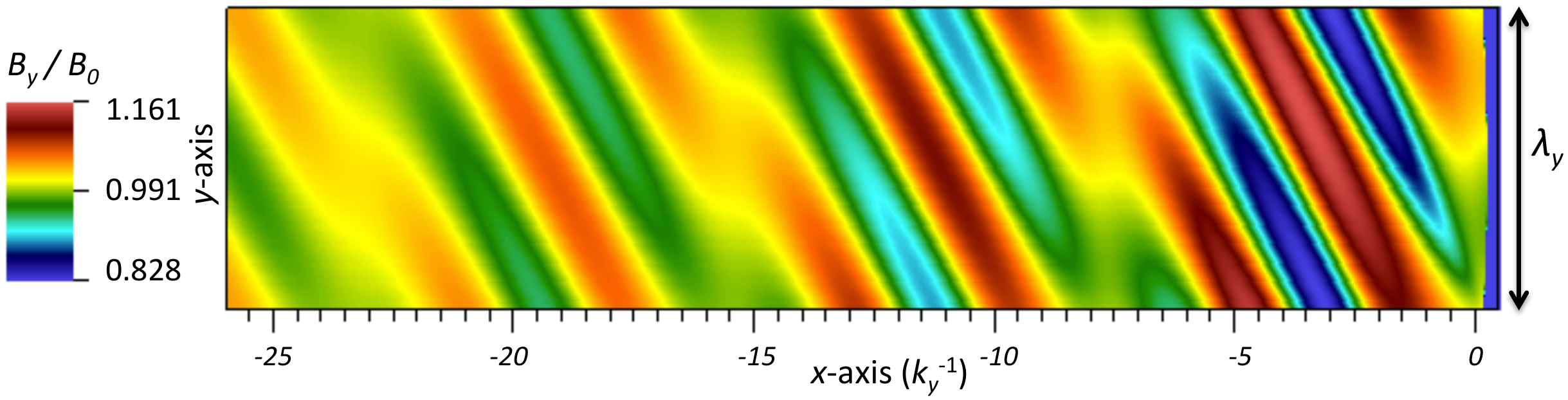}
\caption{Downstream transverse magnetic field (in units of the initial downstream magnetic field $B_0=B_2$) of a simulation of setup 3, corresponding to the interaction of a fast magnetosonic mode with a sub-relativistic
  shock front, at small wave amplitude $\delta \rho / \rho=3\%$ with
  a wavevector close to the resonance, at late times. The color map was truncated
  to enhance the downstream structures.}
  \label{fig:snapshot_ms_s1e-4}
\end{figure*}

The analytical study of LRG16 was conducted in the ultra-relativistic
limit, to simplify the algebra, but there is no obvious physical
reason why resonant corrugation should be a feature of relativistic
shocks alone. In this section, we are thus interested in
sub-relativistic shock velocities and present the results of simulations run for setup 3; whose parameters are close to what one can encounter in supernovae remnants (see Table~\ref{tab:setups}).

\begin{figure}
\center
\includegraphics[scale=1.]{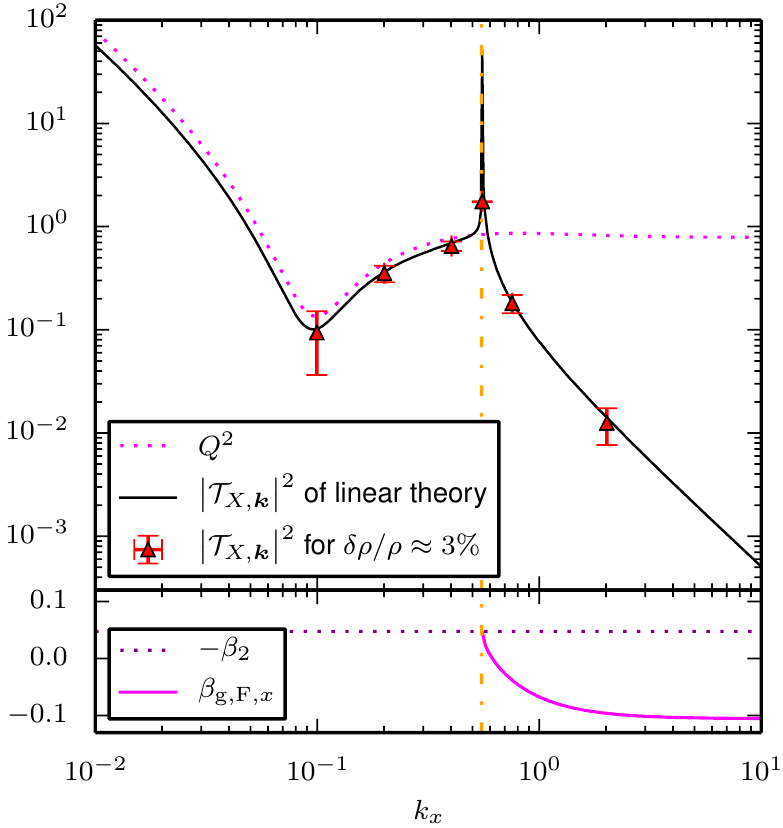}
\caption{Transfer function and group velocity of the outgoing fast mode
  for an incoming fast wave of setup 3 for $k_y=1$. Notations similar
  to Figs.~\ref{plot:tf_e} and \ref{plot:tf_ms}. }
\label{plot:ms_s1e-4}
\end{figure}

Fig.~\ref{fig:snapshot_ms_s1e-4} presents a snapshot from such
simulations and is rather similar to
Fig.~\ref{fig:snapshot_e_a13},~\ref{fig:snapshot_e_a80} and \ref{fig:snapshot_ms}, with the
transmission of modes and their subsequent evolution downstream of the
shock.

Fig.~\ref{plot:ms_s1e-4} presents the corresponding transfer function. As in Fig.~\ref{plot:tf_e} and
\ref{plot:tf_ms}, the solid curve indicates the predictions of linear
theory, borrowed from LRG16 and adapted to the conditions of a
sub-relativistic shock. The theory still predicts a resonant response
when the outgoing fast magnetosonic mode surfs on the shock, as
indicated by the lower panel. This resonance is clearly recovered in
the MHD simulations, at least for small amplitude waves.

Note the dotted line in the upper panel: as before, it represents the quantity $Q^2$, which is related to the perturbation of the incoming energy-momentum in
units of the perturbation amplitude $\delta\psi$. In the present case,
however, this value depends strongly on the incoming wavenumber, in particular,
$Q\,\gg\,1$ for $k_x\,\ll\,k_y$. This behaviour leads to a
large perturbation of the incoming energy-momentum even though
$\delta\rho/\rho\,\ll\,1$, hence to a large corrugation at small
$k_x$. This peculiar $k_x$-dependence of the perturbation appears to be a
feature of fast modes at a low beta-parameter of the plasma rather than a feature of the
sub-relativistic regime.

\section{Summary and discussion}
\label{sec:4}

 \begin{figure*}
\center
\includegraphics[scale=0.63]{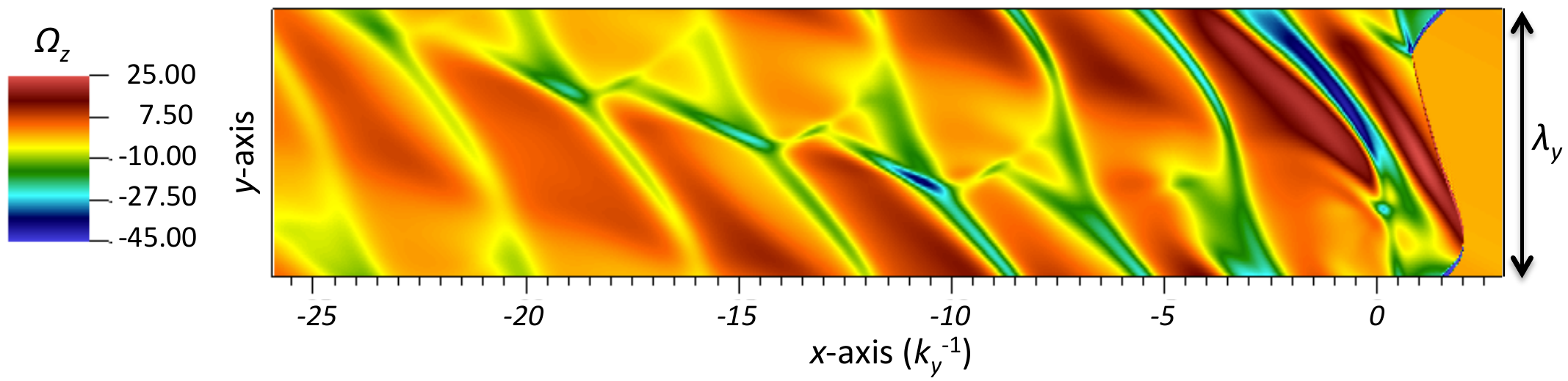}
\caption{Non-vanishing component of the relativistic vorticity field $\Omega^i=\epsilon^{i j}_{\ \ k}\partial_j(w\Gamma \beta^k/\rho)$ in units of $c^2 k_y$ for the same snapshot as in Fig.~\ref{fig:snapshot_e_a80} (incoming entropy wave of amplitude $\delta \rho / \rho\approx0.8$.)}  
\label{fig:vort_e_a80}
\end{figure*}
In this paper, we have discussed 2D SRMHD simulations of a fast
perpendicular shock corrugated by upstream sinusoidal entropy or
magnetosonic waves, in both the relativistic and sub-relativistic flow
velocity regimes. We have measured the transfer function which relates
the amplitude of the corrugation $\Delta X_k$ to that of the incoming
wave $\delta \psi$ and compared it to the predictions of the recent
linear model described in LRG16. The main result of the present study
is that we confirm the existence of a resonant response of the
corrugation, in both the relativistic and the sub-relativistic
regimes, when the fast magnetosonic mode that is produced downstream
of the shock travels at a velocity comparable to that of the shock
front. The interpretation of this resonance is as follows (see LRG16):
as the incoming upstream mode interacts with the shock, it is
transmitted into downstream MHD modes, including one fast magnetosonic
mode which travels possibly as fast as the shock; if this resonance is
satisfied, this mode surfs on and communicates its energy to the shock front,
leading to a large -- possibly formally infinite in the linear theory
-- response of the corrugation pattern. This resonance appears
universal in the sense that, for any setup, for any incoming
perturbation, at any perpendicular wavenumber, there exists at least one value
of the longitudinal wavenumber which satisfies the resonance
criterion.

It may be worth noting that the problem that we discuss here differs
from the Richtmyer-Meshkov instability (RMI)
\citep{Richtmyer1960,Meshkov1969,Brouillette2002,Delmont2009,Nishihara2010}. 
The standard RMI
corresponds to the growth of the perturbation on an
oblique or corrugated interface separating two fluids
after it has encountered a planar shock wave, whereas we have
studied the response of a flat shock front to its
interaction with a compressive perturbation described by a plane
wave. Some authors have
  recently extended the study of the RMI to the case in which the
  shock interacts with a continuous interface separating states of
  different densities with a given contrast
  $A=(\rho_2-\rho_1)/(\rho_2+\rho_1)$ over a finite length scale
  $\lambda$, see e.g. \citet{Brouillette1994}. The problem that we
  have addressed then resembles somewhat this latter configuration,
  since the entropy mode induces a change in density, from
  $\rho_1=\rho-\delta\rho$ to $\rho_2=\rho+\delta\rho$ over a length
  scale $\pi/k_x$, although it repeats this inversion an indefinite
  amount of time, while in the above problem, there is only one
  interface. The previous authors observe that the
growth rate of the RMI falls with increasing $\lambda$, and
decreases with decreasing $A$ as the original RMI. \citet{Zou2017},
on the other hand, examined the case of a rippled shock interacting
with a flat interface and found growth rates much
smaller than those of the "standard" RMI. Both studies however
report hydrodynamic experiments only and the picture probably depends on the
orientation and strength of the magnetic field in the MHD case; parallel shocks
for example, prevent the deposition of vorticity at the interface 
\citep[e.g.]{Sano2013}. In brief, the situation we simulated is quite 
different, we considered waves instead of a unique interface and even if 
triggered, it would seem that the growth rate of the instability is too small 
to be observed within the crossing time of the simulated downstream medium (see
Fig.~\ref{fig:vort_e_a80} for the relativistic vorticity field \citep[e.g.]{2007ApJ...671.1858S} corresponding to the snapshot 
of Fig.\ref{fig:snapshot_e_a80}).

The resonance which we observe is more evocative of the ``spontaneous emission 
of acoustic modes'' described by \citet{DIakov1958}. This author considered the 
interaction of a mode originating from downstream and interacting with a purely 
hydrodynamic shock; he showed that this mode was reflected into the downstream 
medium with a reflection coefficient that could formally become infinite, 
whence the possibility of mode emission in the absence of a perturbation. This 
spontaneous emission has been discussed by a number of authors since then, see 
in particular \citet{Kontorovich1958,Kontorovich1959}, \citet{Landau1987} for 
more qualitative explanations, see \citet{Bates2000} and \citet{Stone1995} for 
numerical illustrations. In our framework, the incoming mode from upstream is 
transmitted through the shock into downstream outgoing modes; the amplitude of 
these modes, just as the amplitude of the shock corrugation, can be obtained 
through the inversion of the response matrix of the linear system. Zeros in the 
determinant of this system then lead to infinitely large responses,
or spontaneous emission. We observe here that this spontaneous emission can (at 
least) take place for some specific values of the longitudinal wavenumber, when 
the outgoing mode surfs on the shock; to our knowledge, this had not been 
noticed before.

For simulations of large amplitude incident waves, the resonance is
partly smoothed out due to some non-linear resonance broadening. At
wave amplitudes outside the realm of linear theory,
i.e. $\delta\psi\,\sim\,\mathcal O(1)$, we observe $k \delta
X_k\,\approx\,1$, whereas $k  \Delta X_k/ \delta \psi >1$ at
$\delta\psi\,\ll\,1$. One caveat is that we have modelled the large
amplitude waves with linear eigenmodes, which are no longer true
eigenmodes of the system of MHD equations; these waves thus tend to
decay into other modes before they enter the shock. It would prove
interesting to conduct simulations with exact non-linear (simple wave)
solutions of the MHD equations. 

In principle, the above resonant response of shock corrugation to
incoming perturbations may find various astrophysical applications,
since the resulting turbulence may have a number of phenomenological
consequences, see e.g. the discussion in Sec.~\ref{sec:introd}. As a
next step in such direction, it would prove useful to conduct
large-scale, high resolution simulations of the interaction of a shock
front with a well developed spectrum of turbulence (as compared to
the present case of a harmonic wave), making sure that the resolution
in $k-$space is sufficient to probe the effect of the resonance. It
would also be highly beneficial to conduct test-particle simulations,
or even hybrid Particle-in-Cell/MHD simulations, in order to study how
the corrugation pattern influences the acceleration process at shock
waves, and how the accelerated particles themselves can induce
corrugation through the instabilities that they develop in the shock
precursor. Incidentally, we note that in a recent paper, \citet{vanMarle2017}
precisely observe the development of an unstable corrugated
configuration, triggered by the interplay of the injection mechanism
with the seeding of turbulence upstream of a corrugated shock.

\section*{Acknowledgements}
We thank the anonymous referee for his/her useful comments.

This work has been financially supported by the ANR-14-CE33-0019 MACH project. 
CD and FC acknowledge the financial support from the UnivEarthS Labex program 
of Sorbonne Paris Cit\'e (ANR-10-LABX-0023 and ANR-11-IDEX-0005-02).  
This work was granted access to HPC resources of CINES under the allocation 
A0020410126 made by GENCI (Grand Equipement National de
Calcul Intensif). \smallskip




\bibliographystyle{mnras}
\bibliography{refs} 



\appendix
\onecolumn
\section{Linear SRMHD eigenmodes}
\label{appendix:modes} 

In MHD, an infinite homogeneous system initially at stationary
equilibrium can develop 7 linearly independent wave modes:
\begin{itemize}
\item one entropy wave (index $_E$): perturbation in the density field only;
\item two Alfv\'en waves:  incompressible and transverse modes; 
\item two slow magnetosonic modes;
\item two fast magnetosonic modes (index $_\mathrm{F}$). 
\end{itemize}
For our purposes, the structure of the perturbations
$\mathrm{\delta \xi}_{\boldsymbol{k}}\equiv (\delta \rho, \delta
p_\mathrm{th}, \delta \beta_{x}, \delta \beta_{y}, \delta B_{x}, \delta
B_{y}) $ of entropy and fast magnetosonic modes in a plasma drifting
at velocity $\beta_0=-\left(1-1/{\Gamma_0}^2\right)^{1/2}$ in the
$x$-direction, initially characterized by the equilibrium $(\rho_0,
p_{\mathrm{th}, 0}, \beta_0, 0, 0, B_0 )$ read:
\begin{align}
&\mathrm{\delta \xi}_{\boldsymbol{k}, \mathrm{E}}= \left(\rho_0, 0, 0, 0, 0, 0 \right) \delta\psi_{\boldsymbol{k}, \mathrm{E}},
\label{eq:perturb_E}\\
&\mathrm{\delta \xi}_{\boldsymbol{k}, \mathrm{F}}=
\left(\frac{{\omega'_{F}}^2-{\beta_{\rm A}}^2 {k'}^2 c^2}{{\beta_{\rm s}}^2
  \left(1-{\beta_{\rm A}}^2 \right){k'_x}^2 c^2}\rho_0,
\frac{{\omega'_{\rm F}}^2-{\beta_{\rm A}}^2 {k'}^2c^2}{{k'_x c^2}^2}W_0,
\frac{\omega'_{\rm F}}{{\Gamma_0}^2 k'_x c}, \frac{{\beta_{\rm s}}^2 \omega'_{\rm F}
  k'_y c}{\Gamma_0 \left({\omega'_{\rm F}}^2-{\beta_{\rm s}}^2 {k'_y}^2c^2\right)},
-\frac{k'_y }{k'_x }\frac{B_0}{\Gamma_0}, B_0 \left( 1+ \beta_0
\frac{\omega'_{\rm F}}{{k'_x c}} \right) \right)
\delta\psi_{\boldsymbol{k}, \mathrm{F}},
\label{eq:perturb_FM}
\end{align}
where $\delta\psi_{\boldsymbol{k}, \mathrm{E/F}}\equiv \delta\psi
\cos(\boldsymbol{k} \cdot \mathbf{x} - \omega_\mathrm{E/F} t)$ is the
harmonic amplitude and primes denote proper quantities measured in the
rest frame of the plasma. Naturally, the velocity/magnetic
perturbations in the lab-frame are just the Lorentz-transformed plasma
rest-frame perturbations. $\omega'$ is linked to $\mathrm{k'}$
through the dispersion relation
\begin{align}
&\omega'_{E}=0, \label{eq:dispersion_E}\\ &\omega'_{\rm F}=\pm\frac{c}{\sqrt{2}}
  \left\{ {\beta_{\rm F}}^2 {k'}^2+{\beta_{\rm A}}^2 {\beta_{\rm s}}^2{k'_y}^2 +\left[ \left(
    {\beta_{\rm F}}^2 {k'}^2+{\beta_{\rm A}}^2 {\beta_{\rm s}}^2{k'_y}^2 \right)^2 -4 {\beta_{\rm A}}^2
    {\beta_{\rm s}}^2{k'_y}^2 {k'}^2 \right]^{1/2} \right\}^{1/2},
\label{eq:dispersion_F}
\end{align}
which involves the characteristic speeds, namely: the sound speed,
${\beta_{\rm s}}\,\equiv\,\left(\gamma_\mathrm{eq} p_{\mathrm{th},
  0}/w_0\right)^{1/2}$, the Alfv\'en speed, ${\beta_{\rm A}}\,=\,{B_0}/
\left({4\pi\Gamma_0}^2 W_0\right)^{1/2}$ and the fast speed, $\beta_{\rm
  F}=\left({\beta_{\rm A}}^2+{\beta_{\rm s}}^2-{\beta_{\rm A}}^2 {\beta_{\rm
    s}}^2 \right)^{1/2}$.

\section{Enforcing divergence-free magnetic field in SRMHD}
\label{appendix:ct}
 
\begin{figure*}
\begin{tabular}{cc}
\includegraphics[width=0.5\textwidth]{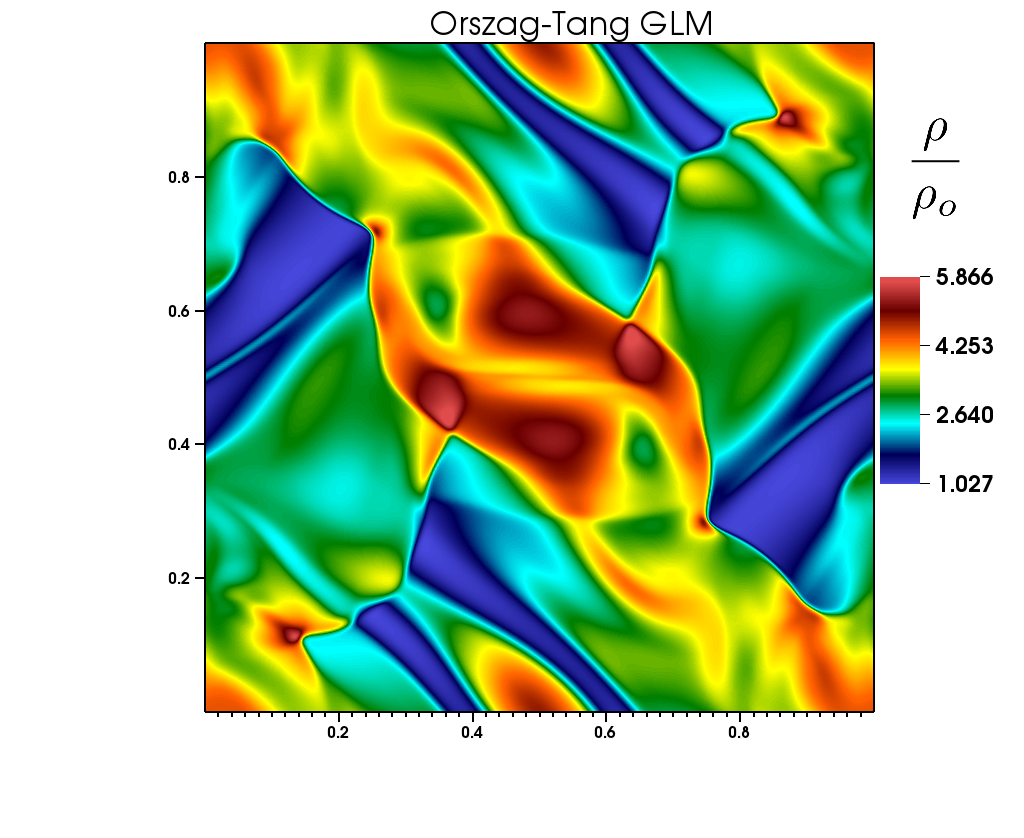} &
\includegraphics[width=0.5\textwidth]{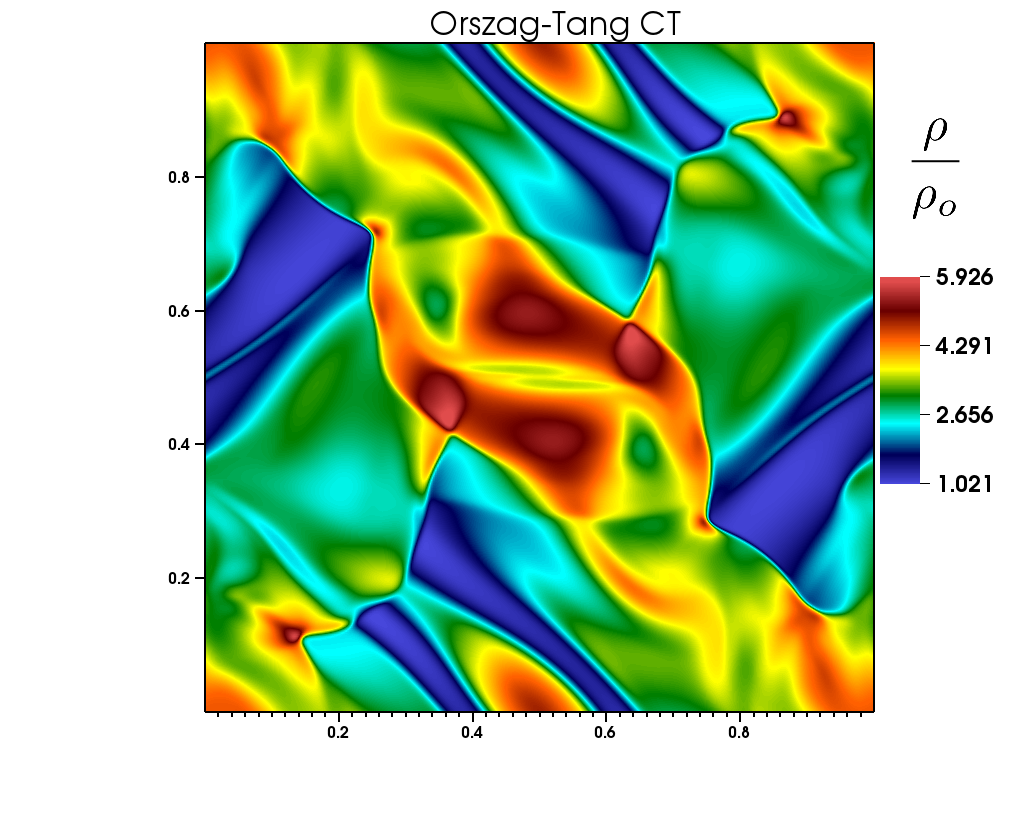} 
\end{tabular}
\caption{Simulations of the relativistic Orszag-Tang test performed with the {\tt MPI-AMRVAC} code using the exact same numerical setup apart for the magnetic divergence cleaning method, namely the GLM ({\bf left}) and the constrained transport ({\bf right}). The proper density of the plasma is displayed at the same point in time, namely $t=1$. The overall density distributions are nearly identical except for some regions as the center of the computational domain or thin mass filaments appearing near the left and right boundaries. These regions actually correspond to areas of the computational domain where the divergence of the magnetic field remains high when using the standard GLM method (see Fig.\ref{Fig:DivB2}).}
\label{Fig:DivB1}
\end{figure*}
\begin{figure*}
\begin{tabular}{cc}
\includegraphics[width=0.5\textwidth]{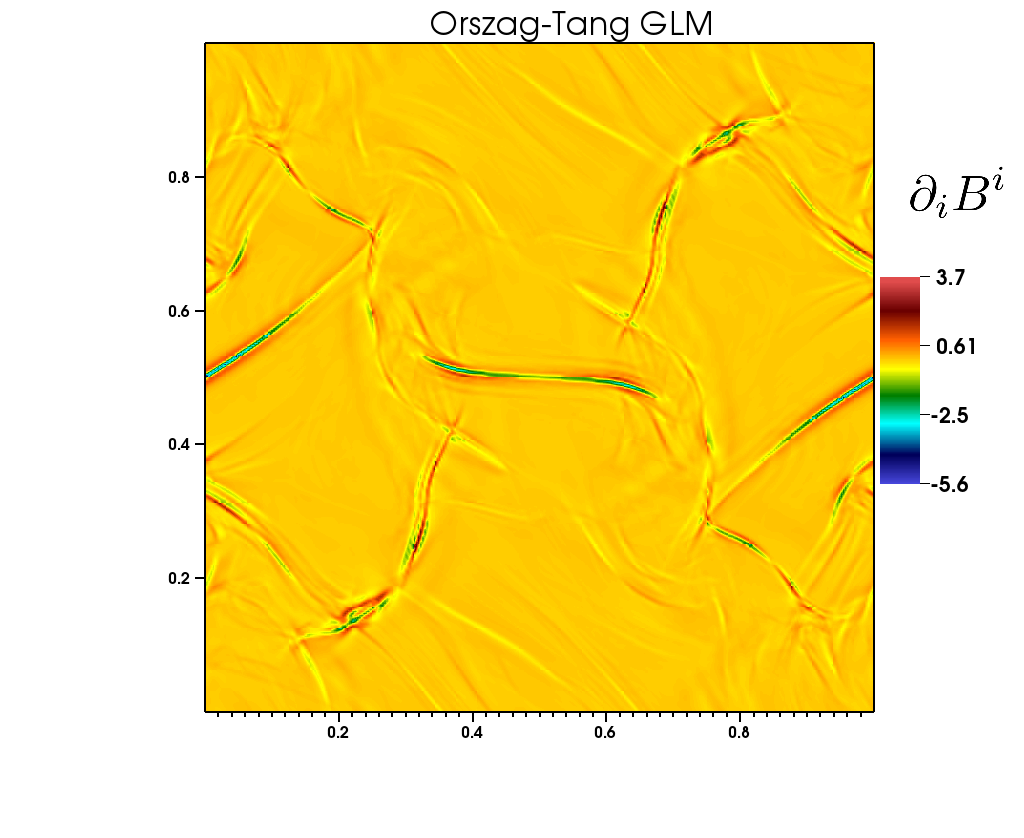} &
\includegraphics[width=0.5\textwidth]{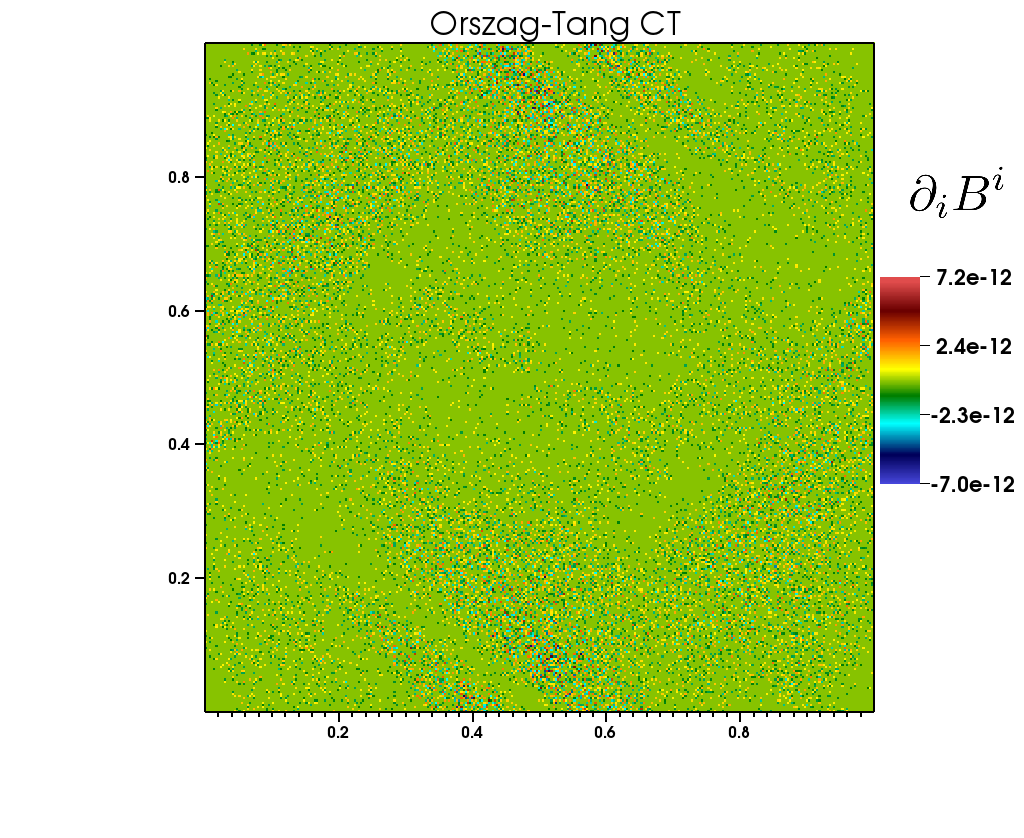} 
\end{tabular}
\caption{Colormaps of the divergence of the magnetic field for two simulations considering the relativistic Orszag-Tang problem with two different magnetic divergence cleaning approaches. The standard hyperbolic divergence cleaning (GLM) does maintain $\partial_iB^i$ to low level except in some narrow regions whereas constrained transport provides vanishing magnetic divergence to machine precision. The regions exhibiting large non-vanishing magnetic divergence actually correspond to areas where the two simulations exhibit small discrepancies regarding plasma quantities.}
\label{Fig:DivB2}
\end{figure*}
All the simulations presented in this paper were performed using a constrained transport (CT) scheme on top of the MHD solver, following the approach of \citet{Balsara1999}. The {\tt MPI-AMRVAC} code also hosts several schemes aiming at cleaning the divergence of the magnetic field. Among these schemes, one of the most efficient one is the hyperbolic divergence cleaning (GLM) developed by \citet{Dedner02}. We have decided to implement the constrained transport method inside the {\tt MPI-AMRVAC} code since considering SRMHD waves requires to maintain the divergence of the magnetic field to zero at machine precision. Such statement is especially important when considering incoming waves onto a shock discontinuity.\\
In order to illustrate the impact of the magnetic divergence cleaning methods in SRMHD simulations, we present results of simulations dealing with the famous Orszag-Tang vortex test \citep{Orszag79}. The relativistic version of this test has been presented in various studies (e.g. \citet{Beckwith11} and references therein). The initial conditions of the simulation are $\rho=25/9$, $P=5/3$ while velocity and magnetic field stand as
\begin{eqnarray}
\beta_x &=& -\beta_o\sin(2\pi y) \nonumber \\
\beta_y &=& \beta_o\sin(2\pi x) \nonumber \\
B_x &=& -\sin (2\pi y) \nonumber \\
B_y &=& \sin (4\pi x) \nonumber 
\end{eqnarray}
where $\beta_o=0.5$ is the maximal velocity of the fluid. The computational domain ranges from zero to unity in both $x$ and $y$ directions while having a grid resolution of $320\times 320$ cells. The simulation has been performed using a TVDLF solver coupled to a minmod slope limiter. Let us also mention that the all boundaries are periodic. 

We have displayed in Fig.\ref{Fig:DivB1} the proper density distribution of the plasma at identical time ($t=1$) for two simulations using different divergence cleaning methods, namely GLM and CT. The two distributions are globally very similar apart in some regions as for instance in the center of the computational domain. Indeed in this region, the GLM simulation has led to a smoother variation of the density compared to the one obtained using the CT algorithm. The same statement actually holds for the magnetic energy density. We can also mention that density filaments appearing in the left and right low density regions are thicker in the GLM simulation than in the CT one. Since both simulations use the very same setup apart from the divergence cleaning approaches, it is likely that these (small) discrepancies stems from local non-zero magnetic divergence. In Fig.\ref{Fig:DivB2} we have displayed the colormaps of $\partial_iB^i$ for the two aforementioned simulations. We then clearly see that large non-vanishing magnetic divergence occurs in zones where the two simulations exhibit differences.

Relativistic simulations of astrophysical shocks deal with fluid velocities very close to the speed of light. Applying an efficient GLM approach to this kind of simulation may become difficult  as the relaxation velocity of magnetic monopoles may not catch up with the fluid evolution, leading to unphysical errors. Among the various magnetic divergence cleaning algorithms published in the literature, we choose to employ the flux constrained transport approach as it provides an efficient way to get rid of magnetic monopoles while preventing any overestimation of the corrugation of the shock \cite[see also the discussion in][]{Toth00}.

\bsp	
\label{lastpage}
\end{document}